\documentclass[%
superscriptaddress,
reprint,
showpacs,
amsmath,amssymb,
aps,
linenumbers,
pre,
]{revtex4-1}

\usepackage{graphicx}% Include figure files
\usepackage{dcolumn}% Align table columns on decimal point
\usepackage{bm}% bold math
\usepackage{color}
\begin {document}
%section {title & abstract}
%subsection {title}
%\preprint{APS/123-QED}

\title{Ultraslow Convergence to Ergodicity in Transient Subdiffusion}
% Force line breaks with \\

\author{Tomoshige Miyaguchi}
\email{tomo@a-phys.eng.osaka-cu.ac.jp}
\affiliation{%
  Department of Applied Physics, Osaka City University, Osaka 558-8585, Japan,
}
\author{Takuma Akimoto}
\affiliation{%
  Department of Mechanical Engineering, Keio University, Yokohama 223-8522, Japan
}%


\date{\today}% It is always \today, today,

%subsection {abstract}

\begin{abstract}
  We investigate continuous time random walks with truncated
  $\alpha$-stable trapping times. We prove distributional ergodicity for a
  class of observables; namely, the time-averaged observables follow the
  probability density function called the Mittag--Leffler
  distribution. This distributional ergodic behavior persists for a long
  time, and thus the convergence to the ordinary ergodicity is considerably
  slower than in the case in which the trapping-time distribution is given
  by common distributions. We also find a crossover from the distributional
  ergodic behavior to the ordinary ergodic behavior.
\end{abstract}

\maketitle

%\tableofcontents

%section {Introduction}
%subsection {background}

The ergodic theorem ensures that time averages of observables converge to
their ensemble averages as the averaging time tends to infinity. On the
other hand, a distributional ergodic theorem states that the probability
density functions (PDFs) of time averages converge to the Mittag--Leffler
(ML) distribution. This property is called infinite ergodicity in dynamical
system theory, because it is associated with infinite invariant measures
\cite{aaronson97, *akimoto10}. Furthermore, in recent years, the
distributional ergodic property has been found for some observables in
stochastic models such as continuous time random walks (CTRWs) \cite{he08,
  *lubelski08, *miyaguchi11b}. For example, the time-averaged mean square
displacement (TAMSD) [Eq.~(\ref{e.tamsd.1})] for CTRWs is a random variable
even in the long measurement time limit and its PDF follows the ML
distribution. It has been pointed out that this distributional ergodic
behavior is reminiscent of the observations in biological experiments that
showed that TAMSDs of macromolecules are widely distributed depending on
trajectories \cite{golding06, *bronstein09, *wang06, *graneli06}. In
addition to these biological systems, CTRW-type systems are used to explain
a broad range of phenomena such as charge carrier transport in amorphous
materials \cite{scher75}, tracer particle diffusion in an array of
convection rolls \cite{young89}, and human mobility \cite{song10}.

%subsection {goal}
One of the important problems on stochastic models such as CTRWs is to
clarify the condition of the distributional ergodicity. It has already been
known that a few observables including the TAMSD show the distributional
ergodicity in CTRWs. But any general criterion for an observable to satisfy
the distributional ergodicity is still unknown. Another important problem
to elucidate is finite size effects \cite{burov11}. For CTRW-type systems,
a power law trapping-time distribution is usually assumed, and thus rare
events---long-time trappings---characterize the long-time behavior. These
rare events, however, are often limited by finite size effects. For
example, if the random trappings are caused by an energetic effect in
complex energy landscapes, the most stable state has the longest trapping
time, thereby causing a cutoff in the trapping-time distribution. In fact,
for the case of macromolecules in cells, the origin of trappings is
considered to be energetic disorder: strong bindings to the target site,
weak bindings to non-specific sites, and intermediate bindings to sites
that are similar to the target site \cite{saxton07}. Because the binding to
the target site should be most stable with the longest trapping time, there
must be a cutoff \cite{saxton07}. Similarly, if the trappings are due to an
entropic effect such as diffusion in inner degrees of freedom (diffusion on
comb-like structures is a simple example \cite{bouchaud90}; see also
\cite{goychuk02}), the finiteness of the phase space of inner degrees of
freedom results in a cutoff. The CTRWs with such a trapping-time cutoff
show distributional ergodic features for short-time measurements, and
become ergodic in the ordinary sense for long-time measurements. But this
transition from distributional ergodic regime to ordinary ergodic regime
has not been elucidated.

%subsection {method}
In this study, we employ a truncated one-sided stable distribution
\cite{mantegna94} as the trapping-time distribution, and show that the
distributional ergodic behavior persists for a remarkably long time
compared to the case of common distributions with the same mean trapping
time. We also show that the time-averaged quantities for a large class of
observables exhibit the distributional ergodicity. As an example, numerical
simulations for a diffusion coefficient are presented. We use the
exponentially truncated stable distribution (ETSD) proposed in
\cite{koponen95, *nakao00} and the numerical method presented in
\cite{gajda10}. This ETSD is useful for rigorous analysis of transient
behavior, because it is an infinitely divisible distribution
\cite{feller71c17} and thus its convoluted distribution or characteristic
function can be explicitly derived [Eqs.~(\ref{e.prob.na}) and
(\ref{e.prob.1b})].

%section {Truncated One-sided L\'evy Flight}
%subsection {infinite divisible distribution}
{\it Truncated one-sided stable distribution.}---In this study, we
investigate CTRWs on $d$-dimensional hypercubic lattices. The lattice
constant is set to unity, and for simplicity, the jumps are allowed only to
the nearest-neighbor sites without preferences. Let $\mbox{\boldmath
  $r$}(t') \in \mathbb{Z}^d$ be the position of the particle at time
$t'$. Moreover, we assume that the successive trapping times
$\tau_k~(k=1,2,...)$ between jumps are mutually independent and the
trapping-time distribution is the ETSD $P_{\rm TL}(\tau, \lambda)$ defined
by the canonical form of the infinitely divisible distribution
\cite{feller71c17}:%p.~581
%% First, let $\psi (\zeta, \lambda)$ be the logarithm of the characteristic
%% function of $P_{\rm TL}(\tau, \lambda)$:
\begin{eqnarray}
  \label{e.prob.1a}
  %% P_{\rm TL}(\tau, \lambda) &=&
  %% \frac {1}{2 \pi}
  %% \int_{-\infty}^{\infty}
  %% e^{\psi (\zeta, \lambda)} e^{-i \zeta \tau} d\zeta,
  %% \\[0.2cm]
  e^{\psi (\zeta, \lambda)}
  &=&
  \int_{-\infty}^{\infty} P_{\rm TL}(\tau, \lambda) e^{i\zeta \tau} d\tau,
  \\[0.05cm]
  \label{e.cfunc1}
  \psi (\zeta, \lambda)
  &=&
  \int_{-\infty}^{\infty}
  \left(
  e^{i\zeta \tau} - 1
  \right)
  f (\tau, \lambda) d\tau,
\end{eqnarray}
%% where $\psi (\zeta, \lambda)$ is the logarithm of the characteristic function
%% of $P_{\rm TL}(\tau, \lambda)$.
The function $f (\tau, \lambda)$ is defined by
\cite{koponen95, *nakao00}
\begin{eqnarray}
  \label{e.incriment}
  f (\tau, \lambda)
  =
  \left\{
  \begin{array}{ll}
    0, &  ~~~(\tau<0)
    \\[0.1cm]
    \displaystyle 
    - c \, \frac {\tau^{-1-\alpha} e^{-\lambda \tau}}{\Gamma(-\alpha)}, &  ~~~(\tau>0),
  \end{array}
  \right.
\end{eqnarray}
where $\Gamma (x)$ is the gamma function, $c>0$ is a scale factor, and
$\alpha \in (0,1)$ is a constant. The parameter $\lambda \geq 0$
characterizes the exponential cutoff [Eq.~(\ref{e.prob.1b})]. When
$\lambda=0$, $P_{\rm TL} (\tau, \lambda)$ is the one-sided $\alpha$-stable
distribution with a power law tail $P_{\rm TL} (\tau, 0) \sim 1 /
\tau^{1+\alpha}$ as $\tau \to \infty$ \cite{feller71c17}. % p.~568
%subsection {convoluted distribution}
The function $\psi (\zeta, \lambda)$ can be expressed as follows:
\begin{eqnarray}
  \label{e.cfunc2}
  \psi (\zeta, \lambda)
  =
  -c \left[ (\lambda - i \zeta)^{\alpha} - \lambda^{\alpha} \right].
\end{eqnarray}
Hence, we obtain $n \psi (\zeta, \lambda) = \psi (n^{1/\alpha} \zeta,
n^{1/\alpha}\lambda)$, where $n \geq 0$ is an integer. Therefore, if $\tau_k
~(k=1,2,...)$ are mutually independent random variables each following $P_{\rm
  TL}(\tau_k, \lambda)$, then the $n$-times convoluted PDF $P_{\rm TL}^{n}
(\tau, \lambda)$, i.e., the PDF of the summation $T_n=\sum_{k=1}^{n} \tau_k$, is
given by
\begin{eqnarray}
  \label{e.prob.na}
  P_{\rm TL}^{n} (\tau, \lambda)
  =
  n^{- 1/ \alpha} P_{\rm TL} (n^{-1/\alpha} \tau, n^{1/\alpha} \lambda).
\end{eqnarray}
% Thus, $P_{\rm TL}^{n} (\tau, \lambda)$ is expressed by $P_{\rm TL} (\tau,
%% \lambda)$.
This is an important outcome of the infinite divisibility and makes it
possible to analyze transient behavior of CTRWs.  Moreover, from
Eq.~(\ref{e.cfunc2}) and the inverse transform of Eq.~(\ref{e.prob.1a}), we
obtain an explicit form of $P_{\rm TL} (\tau, \lambda)$ through the similar
calculation shown in \cite{feller71c17}: %p.~581
\begin{eqnarray}
  P_{\rm TL} (\tau, \lambda) =
  - \frac {e^{ c \lambda^{\alpha} - \lambda \tau}}{\pi \tau}
  \sum_{k=1}^{\infty} \frac {\Gamma (k \alpha + 1)}{k!}
  \left(-c {\tau}^{-\alpha}\right)^k \sin (\pi k \alpha).
  \nonumber\\[-0.4cm]
  \label{e.prob.1b}
\end{eqnarray}

\begin{figure}[]
  \centerline{\includegraphics[width=7.5cm]{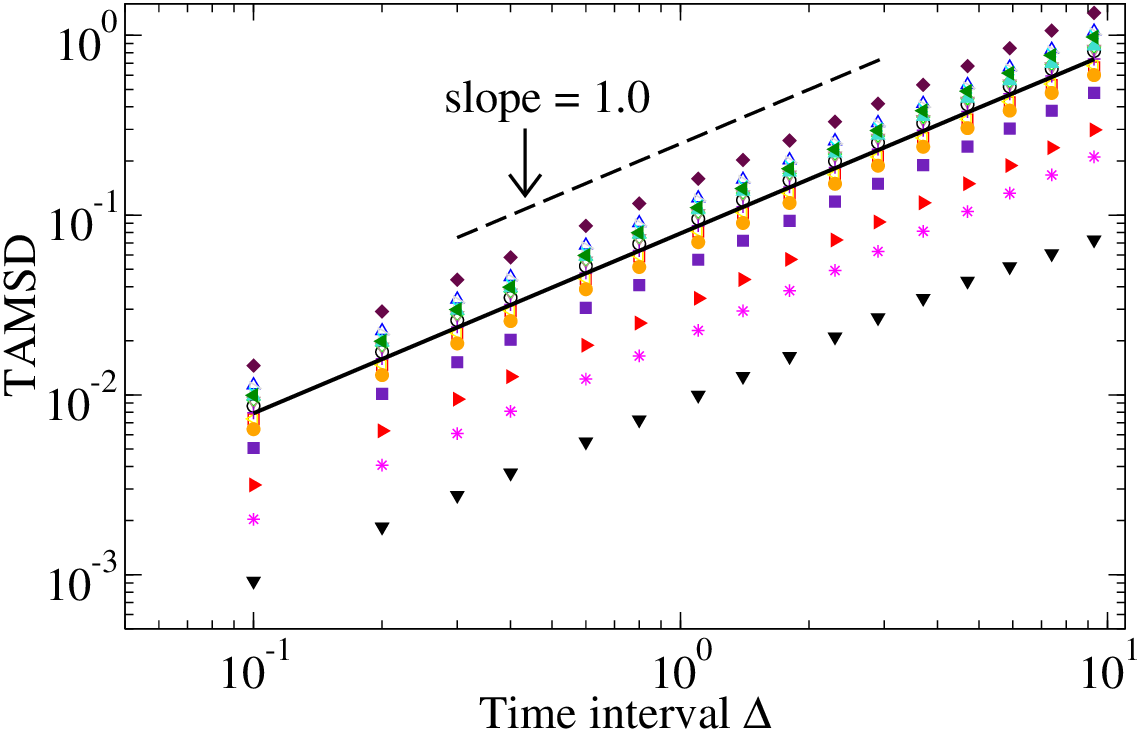}}
  \caption{\label{f.tamsd} (Color online) TAMSD $\overline{(\delta x)^2}
    (\Delta,t)$ vs time interval $\Delta$ (log--log plot) for the
    one-dimensional system ($d=1$). Total measurement time $t$ is set as
    $t=10^{5}$ and other parameters as $\lambda=10^{-5}, ~\alpha=0.75$, and
    $c=1$. The TAMSDs are calculated for 17 different realizations of
    trajectories; a different symbol corresponds to a different
    realization. The solid line is their ensemble average.}
\end{figure}

%section {CTRWs with alpha stable trapping times}
%subsection {hopf's ergodic theorem}
CTRWs {\it with truncated $\alpha$-stable trapping times.}---Now, we
consider the time average of an observable $h(t')$: $\overline{h}_t \equiv
\int_{0}^{t} dt' h(t')/t$, where $t$ is the total measurement time. We
assume that $h(t')$ can be expressed as
\begin{eqnarray}
  \label{e.observable}
  h(t') = \sum_{k=1}^{\infty} H_k \delta(t'-T_k),
\end{eqnarray}
where $T_k > 0~(k=1,2,...)$ is the time when the $k$-th jump occurs, and
$H_k~(k=1,2,...)$ are random variables satisfying $\left\langle H_k
\right\rangle = \left\langle H \right\rangle$ and the ergodicity with
respect to the operational time $k$,
\begin{eqnarray}
  \label{e.H_k}
  \frac {1}{n}\sum_{k=1}^{n} H_k
  \simeq
  \left\langle H \right\rangle,~~~{\rm as}~~~n \to \infty.
\end{eqnarray}
To satisfy Eq.~(\ref{e.H_k}), the correlation function $\left\langle H_k
H_{k+n} \right\rangle - \left\langle H_k \right\rangle \left\langle H_{k+n}
\right\rangle$ should decay more rapidly than $n^{-\gamma}$ with some
constant $\gamma>0$ \cite{bouchaud90,burov10}. % sec. 1.3
It follows from Eqs.~(\ref{e.observable}) and (\ref{e.H_k}) that
\begin{eqnarray}
  \label{e.hopf}
  \overline{h}_t
  =
  %\frac {1}{t}\int_{0}^{t} dt' h(t')
  \frac {1}{t}\sum_{k=1}^{N_t} H_k
  \simeq
  \frac {N_t}{t} \left\langle H \right\rangle,
\end{eqnarray}
for long $t$, where $N_t$ is the number of jumps until time $t$. From this
equation, we find that $\overline{h}_t$ behaves similarly to $N_t$. It is
important that many time-averaged observables for CTRWs can be defined by
Eqs.~(\ref{e.observable}) and (\ref{e.H_k}). For example, the TAMSD,
\begin{eqnarray}
  \label{e.tamsd.1}
  \overline{(\delta \mbox{\boldmath $r$})^2} (\Delta,t)
  \equiv
  \frac {1}{t - \Delta}
  \int_0^{t-\Delta} |\mbox{\boldmath $r$}(t'+\Delta)-\mbox{\boldmath $r$}(t')|^2 dt',~~~
\end{eqnarray}
can be approximately obtained by the time average of $h(t')$ with $H_k$
defined as $H_k \equiv \Delta + 2 \sum_{l=1}^{k-1} (d\mbox{\boldmath $r$}_k
\cdot d\mbox{\boldmath $r$}_l) \theta (\Delta - (T_k - T_l))$, where
$d$-dimensional vector $d \mbox{\boldmath $r$}_k$ is the displacement at
the time $T_k$, and $\theta(t)$ is defined by $\theta(t) = t$ for $t\geq
0$, otherwise $\theta (t) = 0$.  It is easy to see that $\left\langle H_k
\right\rangle = \Delta$ and $\left\langle H_k H_{k+n} \right\rangle -
\left\langle H_k \right\rangle \left\langle H_{k+n} \right\rangle = 0$ for
$n \geq 1$. Using Eq.~(\ref{e.hopf}), we have
%% For CTRWs, we have the following equation for long $t$ \cite{he08,
%%   *miyaguchi10}:
\begin{eqnarray}
  \label{e.tamsd.2}
  \overline{(\delta \mbox{\boldmath $r$})^2} (\Delta,t)
  \simeq
  %%C(\Delta) \frac {N_t}{t},
  \Delta {N_t}/{t}.
\end{eqnarray}
%% The above equation (\ref{e.tamsd.2}) has already derived in somewhat heuristic
%% way \cite{he08,miyaguchi10}.
%%  and $C (\Delta)$ is a constant
%% independent of $t$ and $N_t$. This equation is a result of Hopf's ergodic
%% theorem \cite{aaronson97}. The important point is that the constant $C(\Delta)$
%% is not a random variable, and therefore, the statistical property of TAMSD
%% $\overline{(\delta x)^2} (\Delta,t)$ is determined by that of $N_t$.
%% Furthermore, by the similar calculation shown in \cite{he08, *miyaguchi10}, we
%% have the following equation for $\Delta \ll t$:
%% \begin{eqnarray}
%%   \label{e.tamsd.3}
%%   \langle \overline{(\delta x)^2} (\Delta,t) \rangle
%%   \simeq
%%   %%\Delta \frac {\left\langle N_t\right\rangle}{t}.
%%   \Delta {\left\langle N_t\right\rangle}/{t}.
%% \end{eqnarray}
%% Taking the ensemble average of Eq.~(\ref{e.tamsd.2}), we have $C (\Delta) =
%% \Delta$, that is, TAMSD shows normal diffusion. In contrast, EAMSD shows
%% transient subdiffusion [Eq.~(\ref{e.laplace.1order.b})].
%% From Eq.~(\ref{e.tamsd.2}) and $C (\Delta) = \Delta$,
From Eq.~(\ref{e.tamsd.2}), we obtain a relation between $N_t$ and the diffusion
coefficient of TAMSD as
\begin{eqnarray}
  \label{e.duffusion.const}
  %%D_t \simeq \frac {N_t }{t}.
  D_t \simeq {N_t }/{t}.
\end{eqnarray}
%subsection {fig1.text: tamsd}
In Fig.\ref{f.tamsd}, TAMSDs calculated from 17 different trajectories are
displayed as functions of time interval $\Delta$. This figure shows that
the TAMSD grows linearly with $\Delta$, and the diffusion coefficient $D_t$
is distributed depending on the trajectories.
%subsection {fig1: tamsd}

\begin{figure}[]
  \centerline{\includegraphics[width=7.5cm]{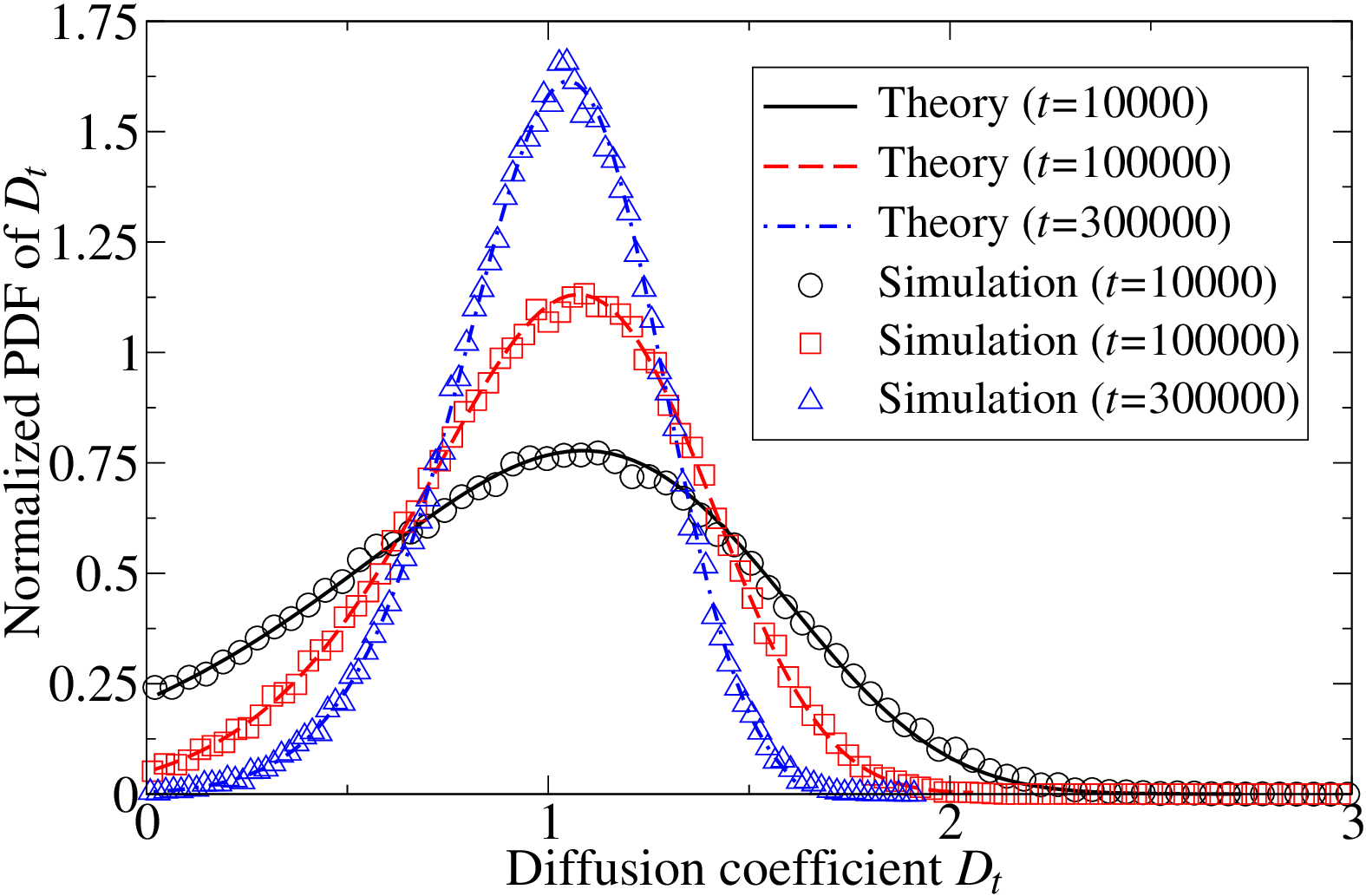}}
  \caption{\label{f.ml} (Color online) The PDF of the diffusion coefficient
    $D_t$ for $d=1$. Each PDF is normalized so that its mean value equals
    unity. $D_t$ is calculated from TAMSD $\overline{(\delta x)^2}
    (\Delta,t)$ by least square fitting over the interval $0 < \Delta <
    10$. The results for three different values of measurement times are
    presented: $t=10^{4}$ (circles), $10^{5}$ (squares), and $3 \times
    10^{5}$ (triangles). The other parameters are set as $\lambda=10^{-5},
    \alpha=0.75$, and $c=1$. The lines correspond to the theoretical
    predictions given by Eq.~(\ref{e.tml}). Note that no adjustable
    parameters were used to obtain these figures.}
\end{figure}

%section {Statistics of diffusion coefficient}
%subsection {Real space analysis}
PDF {\it of time-averaged observables.}---In this section, we derive the
PDF of time-averaged observables $\overline{h}_t$. Because $\overline{h}_t$
and $N_t$ have the same PDF [Eq.~(\ref{e.duffusion.const})], we can study
$N_t$ instead of $\overline{h}_t$.
%% For a trajectory $x(t')$, let $T_n$ be the time when the $n$-th jump occurs.
We have the following relations:
\begin{eqnarray}
  G(n; t) &\equiv& \mathrm{Prob \,} (N_t < n)
  = \mathrm{Prob \,} (T_n > t)
  \nonumber\\[0.10cm]
  &=& \mathrm{Prob} \left( \sum_{k=1}^{n} \tau_k > t \right),
  \label{e.recipro.a}
\end{eqnarray}
where $\mathrm{Prob\,} (\cdot)$ is the probability and $\tau_k$ is the trapping
time between $(k-1)$-th and $k$-th jumps $(k=1,2,...)$. From
Eq.~(\ref{e.recipro.a}), we obtain
\begin{eqnarray}
  G(n; t) &=&
  %% \int_t^{\infty} d\tau ~P_{\rm TL}^{n} (\tau, \lambda)
  %% \nonumber\\[.1cm]
  %% &=&
  %% n^{-1/\alpha} \int_t^{\infty} d\tau ~P_{\rm TL} (n^{-1/\alpha}\tau, n^{1/\alpha}\lambda)
  %% \nonumber\\[.1cm]
  %% &=&
  %% =
  \int_{n^{-1/\alpha}t}^{\infty} d\tau ~P_{\rm TL} (\tau, n^{1/\alpha}\lambda),
%  \nonumber\\[-.3cm]
  \label{e.recipro.b}
\end{eqnarray}
where we have used Eq.~(\ref{e.prob.na}) and the fact that $\tau_k
~(k=1,2,...)$ are mutually independent. Furthermore, we change the
variables from $n$ to $z$ as $ n = t^{\alpha} z$ with $t$ being set.  Then,
by using Eqs.~(\ref{e.prob.1b}), (\ref{e.recipro.a}) and
(\ref{e.recipro.b}), we have
%% \begin{eqnarray}
%%   \mathrm{Pr} \left( \frac {N_t }{ t^{\alpha}} < z \right) 
  %% &=&
  %% \int_{z^{-1/\alpha}}^{\infty} d\tau
  %% ~P_{\rm TL} \left(\tau, t z^{1/\alpha} \lambda \right)
  %% \nonumber\\[.07cm]
%%   \label{e.cumulative.1}
%%   &=&
%%   \int_{0}^{z} \frac {d\tau}{\alpha \tau^{1+1/\alpha}}
%%   ~P_{\rm TL} \left(\frac {1}{\tau^{1/\alpha}}, t z^{1/\alpha} \lambda \right).~~~~~
%% \end{eqnarray}
%% By using Eqs.~(\ref{e.prob.1b}) and (\ref{e.cumulative.1}), we obtain
\begin{eqnarray}
  \mathrm{Prob} \left( \frac {N_t }{ t^{\alpha}} < z \right)
  &=&
  -\frac {e^{c(t\lambda)^{\alpha}z}}{\alpha \pi}
  \sum_{k=1}^{\infty}
  \frac {\Gamma(k \alpha + 1)}{k!k}
  \nonumber \\[0.07cm]
  \label{e.cumulative.2}
  &&\times 
  (-cz)^{k}
  \sin(\pi k \alpha) a_k (t),~~~~~
\end{eqnarray}
where $a_k (t)~(k=1,2,...)$ is defined by
%% \begin{eqnarray}
%%   \label{e.cumulative.a_k}
$a_k (t) \equiv \int_{0}^{1} d\tau e^{-t \lambda \tau^{-1/(\alpha k)}}.$
%% \end{eqnarray}
Differentiating Eq.~(\ref{e.cumulative.2}) with respect to $z$, we have the PDF
of $z=N_t/t^{\alpha}$:
\begin{eqnarray}
  f_{\lambda} (z; t)
  &=&
  -\frac {e^{c(t\lambda)^{\alpha}z}}{\alpha \pi}
  \sum_{k=1}^{\infty}
  \frac {\Gamma(k \alpha + 1)}{k!} (-c)^{k}
  \nonumber \\[0.1cm]
  \label{e.tml}
  &&\times 
  \left[ \frac {c(t \lambda)^{\alpha} z}{k} + 1\right] z^{k-1}
  \sin(\pi k \alpha) a_k(t).~~~~~
\end{eqnarray}
Because of Eq.~(\ref{e.hopf}), Eq.~(\ref{e.tml}) is the PDF of the
time-averaged observables $\overline{h}_t t^{1-\alpha}/\left\langle H
\right\rangle$ including the diffusion constant $D_tt^{1-\alpha}$
[Eq.~(\ref{e.duffusion.const})] as a special case. When $\lambda=0$, the PDF
$f_{0}(z)$, which is the ML distribution \cite{aaronson97, akimoto10}, is
time-independent. Namely, the time-averaged observables are random variables
even in the limit $t \to \infty$; this property is called the distributional
ergodicity. On the other hand, when $\lambda > 0$, the PDF tends to a delta
function. Thus, the time-averages converge to constant values as is expected
from the ordinary ergodicity.
%subsection {fig2.text: ml pdf}
The PDF of $D_t$ is shown in Fig.~\ref{f.ml} for three different measurement
times $t$. It is clear that the PDF becomes narrower for a longer $t$. The
analytical result given by Eq.~(\ref{e.tml}) is also illustrated by the lines.
%% and they show good agreement with the numerical results.

%subsection {fig2: ml pdf}
%\clearpage
%subsection {Laplace space analysis}

\begin{figure}[]
  \centerline{\includegraphics[width=7.5cm]{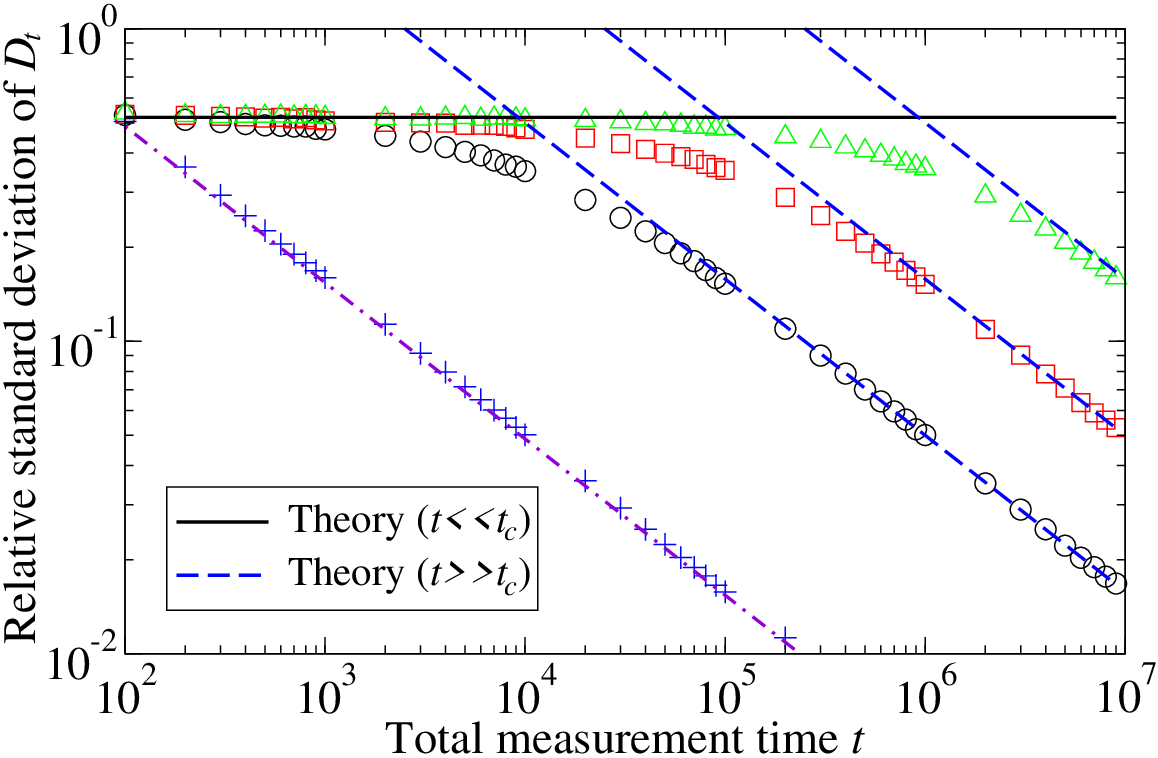}}
  \caption{\label{f.relative_sd} (Color online) RSD $\sqrt{\left\langle
      D_t^2 \right\rangle_c} / \left\langle D_t \right\rangle$ vs total
    measurement time $t$ for $d=1$. $D_t$ is calculated from the TAMSD
    $\overline{(\delta x)^2} (\Delta,t)$ by least square fitting over the
    interval $0 < \Delta < 1$. Three different values of $\lambda$ are
    used: $\lambda=10^{-4}$ (circles), $10^{-5}$ (squares), and $10^{-6}$
    (triangles), and $\alpha$ and $c$ are set as $\alpha=0.75$ and $c=1$,
    respectively. The lines correspond to the theoretical prediction given
    by Eq.~(\ref{e.relative.sd}); the solid line is the result for short
    time scales $t\ll t_c$, whereas the dashed lines are for long time
    scales $t \gg t_c$. The intersections of the solid and dashed lines
    correspond to the crossover times $t_c$ given by Eq.~(\ref{e.tc}). The
    pluses are the RSD for the case of the exponential distribution
    $P(\tau) = \exp(\tau / \left\langle \tau \right\rangle) / \left\langle
    \tau \right\rangle$ with the same mean trapping time as the ETSD with
    $\lambda = 10^{-6}$ (triangles): $\left\langle \tau \right\rangle = c
    \lambda^{\alpha-1} \alpha$. The dot-dashed line is a theoretical
    prediction for the exponential distribution: $R(t)=(c\alpha
    \lambda^{\alpha-1}/t)^{1/2}$.}
\end{figure}

{\it Relative standard deviation.}---Next, in order to quantify a deviation from
the ordinary ergodicity, we study a relative standard deviation (RSD) of
time-averaged observables $R(t)=\sqrt{\langle (\overline{h}_t)^2 \rangle_c} /
\left\langle \overline{h}_t \right\rangle$, where $\left\langle \cdot
\right\rangle$ is the ensemble average over trajectories and $\left\langle \cdot
\right\rangle_c$ is the corresponding cumulant.
%% The RSD $R(t)$ is closely related to the ergodicity
%% breaking parameter \cite{he08}.
If $R(t) \approx 0$, the system can be considered to be ergodic in the ordinary
sense, whereas if $R(t) > 0$, the system is not ergodic. To derive an analytical
expression for $R(t)$, we take the Laplace transform of Eq~(\ref{e.recipro.b})
with respect to $t$:
\begin{eqnarray}
  \label{e.laplace.1}
  \tilde{G}(n; s)
  %% &\equiv&
  %% \int_{0}^{\infty} dt e^{-ts}
  %% \int_{n^{-1/\alpha}t}^{\infty} d\tau ~P_{\rm TL} (\tau, n^{1/\alpha}\lambda) 
  %% \nonumber\\[.15cm]
  %% &=&
  =
  \frac {1 - e^{-nc[(\lambda + s)^{\alpha} - \lambda^{\alpha}]}}{s},
\end{eqnarray}
where we have defined $\tilde{G}(n; s)$ as $\tilde{G}(n; s) \equiv
\int_{0}^{\infty} dt e^{-ts} G (n; t)$ and used Eq.~(\ref{e.cfunc2}).
%with $\zeta = i s$.
Next, we define a function $g (n;s)$ by $g (n;s):= \tilde{G} (n+1; s) -
\tilde{G} (n; s)$. Then, by taking a (discrete) Laplace transform with respect
to $n$, $\sum_{n=0}^{\infty} e^{-n\nu} g(n;s) \equiv \tilde{g}(\nu; s)$, we have
%% \begin{eqnarray}
%%   \label{e.laplace.2}
%%   \tilde{g}(\nu; s) =
%%   \frac {1}{s}
%%   \frac
%%   {1- \exp \left( -c[(\lambda + s)^{\alpha} - \lambda^{\alpha}] \right)}
%%   {1- \exp \left( - \nu - c[(\lambda + s)^{\alpha} - \lambda^{\alpha}] \right)}.
%% \end{eqnarray}
\begin{eqnarray}
  \label{e.laplace.3}
  \tilde{g}(\nu; s)
  \simeq
  %% \frac {1}{s}
  %% \frac
  %% {c[(\lambda + s)^{\alpha} - \lambda^{\alpha}]}
  %% {\nu + c[(\lambda + s)^{\alpha} - \lambda^{\alpha}]}
  %% \nonumber\\[.2cm]
  %% &=& 
  %% \frac {1}{s}
  %% \times
  %% \frac
  %% {1}
  %% {1 + \frac {\nu}{c[(\lambda + s)^{\alpha} - \lambda^{\alpha}]}}
  %% \nonumber\\[.2cm]
  %% &=&
  \frac {1}{s}
  \sum_{k=0}^{\infty}
  \left(- \frac {\nu}{c}\right)^k
  \left[(\lambda + s)^{\alpha} - \lambda^{\alpha}\right]^{-k},
\end{eqnarray}
where we used an approximation by assuming $s, \lambda \ll 1$.
%subsection {first moments}
From Eq.~(\ref{e.laplace.3}), we can derive arbitrary order of moments of
$N_t$. 
%% the Laplace transform $\mathcal{L}[\left\langle N_t
%% \right\rangle](s)$ of the first moment $\left\langle N_t\right\rangle$ is given
%% by
%% \begin{eqnarray}
%%   \label{e.laplace.1order.a}
%%   \mathcal{L}[\left\langle N_t \right\rangle](s)
%%   %% &\simeq&
%%   %% \frac {1}{cs}
%%   %% \frac {1}{ (\lambda + s)^{\alpha} - \lambda^{\alpha} }
%%   %% \nonumber\\[.2cm]
%%   &\simeq&
%%   \left\{
%%   \begin{array}{ll}
%%     \displaystyle 
%%     \frac {1}{c s^{\alpha+1}},& \hspace*{.1cm} s\gg \lambda \\[.3cm]
%%     \displaystyle 
%%     \frac {1}{c \lambda^{\alpha-1} \alpha s^{2}}
%%     \left[ 1 + (1 - \alpha) \frac {s}{2 \lambda}\right],
%%     & \hspace*{.1cm} s \ll \lambda.~~~~~
%%   \end{array}
%%   \right.
%% \end{eqnarray}
%% In the real space,
For example, the first moment $\left\langle N_t\right\rangle$ is given by
\begin{eqnarray}
  \label{e.laplace.1order.b}
  \left\langle N_t\right\rangle
  \simeq
  \left\{
  \begin{array}{ll}
    \displaystyle 
    \frac {t^{\alpha}}{c \Gamma (\alpha + 1)},& \hspace*{.3cm} t \ll 1/\lambda \\[.25cm]
    \displaystyle 
    \frac {t}{c \lambda^{\alpha-1} \alpha } + \frac {1 - \alpha}{2 c \lambda^{\alpha} \alpha },
    &\hspace*{.3cm} t \gg 1/\lambda.
  \end{array}
  \right.
\end{eqnarray}
The ensemble-averaged mean square displacement (EAMSD) for CTRWs is known
to be proportional to $\left\langle N_t \right\rangle$ \cite{bouchaud90}:
$\left\langle (\delta \mbox{\boldmath $r$})^2 \right\rangle (t) \sim
\left\langle N_t \right\rangle$. Thus, the EAMSD of the present model shows
transient subdiffusion, i.e., subdiffusion for short time scales and normal
diffusion for long timescales \cite{saxton07}.
%subsection {second moments}
%% Similarly, the second moment can be derived as follows:
%% \begin{eqnarray}
%%   \label{e.laplace.2order.b}
%%   \left\langle N_t^{2}\right\rangle
%%   \simeq
%%   \left\{
%%   \begin{array}{ll}
%%     \displaystyle 
%%     \frac {2 t^{2 \alpha}}{c^{2}\Gamma (2 \alpha + 1)},& \hspace*{.3cm} t \ll 1/\lambda \\[.3cm]
%%     \displaystyle 
%%     \frac {2}{c^{2} \lambda^{2\alpha-2} \alpha^{2} }
%%     \left[ \frac {t^{2}}{2} + \frac {1-\alpha}{\lambda}t\right],
%%     & \hspace*{.3cm} t \gg 1/\lambda.
%%   \end{array}
%%   \right.
%% \end{eqnarray}
%subsection {relative standard deviation}
Similarly, the second moment can be derived and we have the RSD for $N_t$ as
follows:
%% From Eqs.~(\ref{e.laplace.1order.b}) and (\ref{e.laplace.2order.b}), we have the
%% RSD for $N_t$ as follows:
\begin{eqnarray}
  \label{e.relative.sd}
  \frac {\sqrt{\left\langle N_t^2 \right\rangle_c}}{  \left\langle N_t \right\rangle}
  \simeq
  \left\{
  \begin{array}{ll}
    %  \displaystyle 
    \sqrt{\frac {2 \Gamma^{2} (\alpha + 1)}{\Gamma (2 \alpha + 1)} - 1 },
    & \hspace*{.3cm} t \ll 1/\lambda \\[.0cm]
    %  \displaystyle
    (1 - \alpha)^{1/2} \lambda^{-1/2} t^{-1/2},
    & \hspace*{.3cm} t \gg 1/\lambda.
  \end{array}
  \right.
\end{eqnarray}
Note that the RSDs for $\overline{h}_t$ and $D_t$ also follow the same
relations, because they differ only in the scale factor [Eqs.~(\ref{e.hopf}) and
(\ref{e.duffusion.const})]. From these results, the crossover time $t_c$ between
the distributional and ordinary ergodic regimes is given by
\begin{eqnarray}
  \label{e.tc}
  t_c
  =
  %\frac {(1-\alpha)}{\frac {2 \Gamma^{2}(\alpha + 1)}{\Gamma (2\alpha + 1) }- 1}
  %\lambda^{-1}.
  \frac {(1-\alpha)}{{2 \Gamma^{2}(\alpha + 1)} / {\Gamma (2\alpha + 1) }- 1} \lambda^{-1}.
\end{eqnarray}
%subsection {fig3.text: rsd}
As shown in Fig.~\ref{f.relative_sd}, the RSD remains almost constant
before the crossover time $t_c$, and starts to decay rapidly after the
crossover.  In Fig.~\ref{f.relative_sd}, the RSD for the exponential
trapping-time distribution which has the same mean trapping time
$\left\langle \tau \right\rangle$ as the ETSD with $\lambda = 10^{-6}$ is
also shown by pluses. It is clear that the RSD for exponential distribution
(pluses) decays much more rapidly than that for the ETSD (triangles).

%subsection {fig3: rsd}
%\clearpage
%section {summary}

{\it Summary.}---In this study, we have investigated the CTRWs with
truncated $\alpha$-stable trapping times. The three main results are as
follows: (i)~We proved the distributional ergodicity for short measurement
times; namely, the time averages of observables behave as random variables
following the ML distribution. Moreover, we derived the PDF at arbitrary
measurement times. It is very interesting to compare this analytical
formula [Eq.~(\ref{e.tml})] with the results for lipid granules reported
recently \cite{jeon11}. We should also note that the limit distributions,
the ML distribution for the case of observables studied in this paper,
depends on the definition of observables \cite{rebenshtok07, *rebenshtok08,
  *akimoto08b}. (ii) We found that the distributional ergodic behavior
persists for a long time. In other words, the convergence to the ordinary
ergodicity is remarkably slow in contrast to the case in which the
trapping-time distribution is given by common distributions such as the
exponential distribution. This indicates that, in real experiments, the
time-averaged quantities could behave as random variables even for
considerably long measurement times. (iii)~We found a crossover from the
distributional ergodicity in the short-time regime to the ordinary
ergoodicity in the long-time regime.
%% Surprisingly, this crossover is a kind of order--disorder phase
%% transition in the limit $\lambda \to 0$ with the RSD $R(t)$ as an order
%% parameter.
Finally, it is worth mentioning that these three main results are valid for
a large class of observables. This implies that it is possible to choose an
observable which is easy to measure experimentally. Then, the system
parameters $\alpha$ and $\lambda$ can be experimentally determined by the
short-- and long--time behavior of the RSD $R(t)$
[Eq.~(\ref{e.relative.sd})], respectively.

%% It is also worth mentioning that this crossover and the slow convergence to
%% ergodicity occur even when the molecule is confined in a finite region
%% \cite{neusius09, *miyaguchi11b}.

%% We thank Y.~Aizawa for helpful discussions and encouragement. This work is
%% partly supported by a Grant-in-Aid for Young Scientists (B) (22740262).

%section {bibliography}

%\bibliography{paper}

\begin{thebibliography}{26}%
\makeatletter
\providecommand \@ifxundefined [1]{%
 \@ifx{#1\undefined}
}%
\providecommand \@ifnum [1]{%
 \ifnum #1\expandafter \@firstoftwo
 \else \expandafter \@secondoftwo
 \fi
}%
\providecommand \@ifx [1]{%
 \ifx #1\expandafter \@firstoftwo
 \else \expandafter \@secondoftwo
 \fi
}%
%
%
\providecommand \bibnamefont  [1]{#1}%
\providecommand \bibfnamefont [1]{#1}%
\providecommand \citenamefont [1]{#1}%
\providecommand \href@noop [0]{\@secondoftwo}%
\providecommand \href [0]{\begingroup \@sanitize@url \@href}%
\providecommand \@href[1]{\@@startlink{#1}\@@href}%
\providecommand \@@href[1]{\endgroup#1\@@endlink}%
\providecommand \@sanitize@url [0]{\catcode `\\12\catcode `\$12\catcode
  `\&12\catcode `\#12\catcode `\^12\catcode `\_12\catcode `\%12\relax}%
\providecommand \@@startlink[1]{}%
\providecommand \@@endlink[0]{}%
\providecommand \url  [0]{\begingroup\@sanitize@url \@url }%
\providecommand \@url [1]{\endgroup\@href {#1}{\urlprefix }}%
\providecommand \urlprefix  [0]{URL }%
%
\providecommand \doibase [0]{http://dx.doi.org/}%
\providecommand \selectlanguage [0]{\@gobble}%
\providecommand \bibinfo  [0]{\@secondoftwo}%
\providecommand \bibfield  [0]{\@secondoftwo}%
%
\providecommand \BibitemOpen [0]{}%
%
%
%
\providecommand \BibitemShut  [1]{\csname bibitem#1\endcsname}%
\let\auto@bib@innerbib\@empty
%</preamble>
\bibitem [{\citenamefont {Aaronson}(1997)}]{aaronson97}%
  \BibitemOpen
  \bibfield  {author} {\bibinfo {author} {\bibfnamefont {J.}~\bibnamefont
  {Aaronson}},\ }\href@noop {} {\emph {\bibinfo {title} {An Introduction to
  Infinite Ergodic Theory}}}\ (\bibinfo  {publisher} {American Mathematical
  Society},\ \bibinfo {address} {Province},\ \bibinfo {year}
  {1997})\BibitemShut {NoStop}%
\bibitem [{\citenamefont {Akimoto}\ and\ \citenamefont
  {Miyaguchi}(2010)}]{akimoto10}%
  \BibitemOpen
  \bibfield  {author} {\bibinfo {author} {\bibfnamefont {T.}~\bibnamefont
  {Akimoto}}\ and\ \bibinfo {author} {\bibfnamefont {T.}~\bibnamefont
  {Miyaguchi}},\ }\href {\doibase 10.1103/PhysRevE.82.030102} {\bibfield
  {journal} {\bibinfo  {journal} {Phys. Rev. E}\ }\textbf {\bibinfo {volume}
  {82}},\ \bibinfo {pages} {030102} (\bibinfo {year} {2010})}\BibitemShut
  {NoStop}%
\bibitem [{\citenamefont {He}\ \emph {et~al.}(2008)\citenamefont {He},
  \citenamefont {Burov}, \citenamefont {Metzler},\ and\ \citenamefont
  {Barkai}}]{he08}%
  \BibitemOpen
  \bibfield  {author} {\bibinfo {author} {\bibfnamefont {Y.}~\bibnamefont
  {He}}, \bibinfo {author} {\bibfnamefont {S.}~\bibnamefont {Burov}}, \bibinfo
  {author} {\bibfnamefont {R.}~\bibnamefont {Metzler}}, \ and\ \bibinfo
  {author} {\bibfnamefont {E.}~\bibnamefont {Barkai}},\ }\href@noop {}
  {\bibfield  {journal} {\bibinfo  {journal} {Phys. Rev. Lett.}\ }\textbf
  {\bibinfo {volume} {101}},\ \bibinfo {pages} {058101} (\bibinfo {year}
  {2008})}\BibitemShut {NoStop}%
\bibitem [{\citenamefont {Lubelski}\ \emph {et~al.}(2008)\citenamefont
  {Lubelski}, \citenamefont {Sokolov},\ and\ \citenamefont
  {Klafter}}]{lubelski08}%
  \BibitemOpen
  \bibfield  {author} {\bibinfo {author} {\bibfnamefont {A.}~\bibnamefont
  {Lubelski}}, \bibinfo {author} {\bibfnamefont {I.~M.}\ \bibnamefont
  {Sokolov}}, \ and\ \bibinfo {author} {\bibfnamefont {J.}~\bibnamefont
  {Klafter}},\ }\href@noop {} {\bibfield  {journal} {\bibinfo  {journal} {Phys.
  Rev. Lett.}\ }\textbf {\bibinfo {volume} {100}},\ \bibinfo {pages} {250602}
  (\bibinfo {year} {2008})}\BibitemShut {NoStop}%
\bibitem [{\citenamefont {Miyaguchi}\ and\ \citenamefont
  {Akimoto}(2011)}]{miyaguchi11b}%
  \BibitemOpen
  \bibfield  {author} {\bibinfo {author} {\bibfnamefont {T.}~\bibnamefont
  {Miyaguchi}}\ and\ \bibinfo {author} {\bibfnamefont {T.}~\bibnamefont
  {Akimoto}},\ }\href {\doibase 10.1103/PhysRevE.83.031926} {\bibfield
  {journal} {\bibinfo  {journal} {Phys. Rev. E}\ }\textbf {\bibinfo {volume}
  {83}},\ \bibinfo {pages} {031926} (\bibinfo {year} {2011})}\BibitemShut
  {NoStop}%
\bibitem [{\citenamefont {Golding}\ and\ \citenamefont
  {Cox}(2006)}]{golding06}%
  \BibitemOpen
  \bibfield  {author} {\bibinfo {author} {\bibfnamefont {I.}~\bibnamefont
  {Golding}}\ and\ \bibinfo {author} {\bibfnamefont {E.~C.}\ \bibnamefont
  {Cox}},\ }\href@noop {} {\bibfield  {journal} {\bibinfo  {journal} {Phys.
  Rev. Lett.}\ }\textbf {\bibinfo {volume} {96}},\ \bibinfo {pages} {098102}
  (\bibinfo {year} {2006})}\BibitemShut {NoStop}%
\bibitem [{\citenamefont {Bronstein}\ \emph {et~al.}(2009)\citenamefont
  {Bronstein}, \citenamefont {Israel}, \citenamefont {Kepten}, \citenamefont
  {Mai}, \citenamefont {{Shav-Tal}}, \citenamefont {Barkai},\ and\
  \citenamefont {Garini}}]{bronstein09}%
  \BibitemOpen
  \bibfield  {author} {\bibinfo {author} {\bibfnamefont {I.}~\bibnamefont
  {Bronstein}}, \bibinfo {author} {\bibfnamefont {Y.}~\bibnamefont {Israel}},
  \bibinfo {author} {\bibfnamefont {E.}~\bibnamefont {Kepten}}, \bibinfo
  {author} {\bibfnamefont {S.}~\bibnamefont {Mai}}, \bibinfo {author}
  {\bibfnamefont {Y.}~\bibnamefont {{Shav-Tal}}}, \bibinfo {author}
  {\bibfnamefont {E.}~\bibnamefont {Barkai}}, \ and\ \bibinfo {author}
  {\bibfnamefont {Y.}~\bibnamefont {Garini}},\ }\href {\doibase
  10.1103/PhysRevLett.103.018102} {\bibfield  {journal} {\bibinfo  {journal}
  {Phys. Rev. Lett.}\ }\textbf {\bibinfo {volume} {103}},\ \bibinfo {pages}
  {018102} (\bibinfo {year} {2009})}\BibitemShut {NoStop}%
\bibitem [{\citenamefont {Wang}\ \emph {et~al.}(2006)\citenamefont {Wang},
  \citenamefont {Austin},\ and\ \citenamefont {Cox}}]{wang06}%
  \BibitemOpen
  \bibfield  {author} {\bibinfo {author} {\bibfnamefont {Y.~M.}\ \bibnamefont
  {Wang}}, \bibinfo {author} {\bibfnamefont {R.~H.}\ \bibnamefont {Austin}}, \
  and\ \bibinfo {author} {\bibfnamefont {E.~C.}\ \bibnamefont {Cox}},\ }\href
  {\doibase 10.1103/PhysRevLett.97.048302} {\bibfield  {journal} {\bibinfo
  {journal} {Phys. Rev. Lett.}\ }\textbf {\bibinfo {volume} {97}},\ \bibinfo
  {pages} {048302} (\bibinfo {year} {2006})}\BibitemShut {NoStop}%
\bibitem [{\citenamefont {Gran\'eli}\ \emph {et~al.}(2006)\citenamefont
  {Gran\'eli}, \citenamefont {Yeykal}, \citenamefont {Robertson},\ and\
  \citenamefont {Greene}}]{graneli06}%
  \BibitemOpen
  \bibfield  {author} {\bibinfo {author} {\bibfnamefont {A.}~\bibnamefont
  {Gran\'eli}}, \bibinfo {author} {\bibfnamefont {C.~C.}\ \bibnamefont
  {Yeykal}}, \bibinfo {author} {\bibfnamefont {R.~B.}\ \bibnamefont
  {Robertson}}, \ and\ \bibinfo {author} {\bibfnamefont {E.~C.}\ \bibnamefont
  {Greene}},\ }\href {\doibase 10.1073/pnas.0508366103} {\bibfield  {journal}
  {\bibinfo  {journal} {Proc. Natl. Acad. Sci. U.S.A}\ }\textbf {\bibinfo
  {volume} {103}},\ \bibinfo {pages} {1221 } (\bibinfo {year}
  {2006})}\BibitemShut {NoStop}%
\bibitem [{\citenamefont {Scher}\ and\ \citenamefont
  {Montroll}(1975)}]{scher75}%
  \BibitemOpen
  \bibfield  {author} {\bibinfo {author} {\bibfnamefont {H.}~\bibnamefont
  {Scher}}\ and\ \bibinfo {author} {\bibfnamefont {E.~W.}\ \bibnamefont
  {Montroll}},\ }\href {\doibase 10.1103/PhysRevB.12.2455} {\bibfield
  {journal} {\bibinfo  {journal} {Phys. Rev. B}\ }\textbf {\bibinfo {volume}
  {12}},\ \bibinfo {pages} {2455} (\bibinfo {year} {1975})}\BibitemShut
  {NoStop}%
\bibitem [{\citenamefont {Young}\ \emph {et~al.}(1989)\citenamefont {Young},
  \citenamefont {Pumir},\ and\ \citenamefont {Pomeau}}]{young89}%
  \BibitemOpen
  \bibfield  {author} {\bibinfo {author} {\bibfnamefont {W.}~\bibnamefont
  {Young}}, \bibinfo {author} {\bibfnamefont {A.}~\bibnamefont {Pumir}}, \ and\
  \bibinfo {author} {\bibfnamefont {Y.}~\bibnamefont {Pomeau}},\ }\href
  {\doibase 10.1063/1.857415} {\bibfield  {journal} {\bibinfo  {journal} {Phys.
  Fluids A}\ }\textbf {\bibinfo {volume} {1}},\ \bibinfo {pages} {462}
  (\bibinfo {year} {1989})}\BibitemShut {NoStop}%
\bibitem [{\citenamefont {Song}\ \emph {et~al.}(2010)\citenamefont {Song},
  \citenamefont {Koren}, \citenamefont {Wang},\ and\ \citenamefont
  {Barabasi}}]{song10}%
  \BibitemOpen
  \bibfield  {author} {\bibinfo {author} {\bibfnamefont {C.}~\bibnamefont
  {Song}}, \bibinfo {author} {\bibfnamefont {T.}~\bibnamefont {Koren}},
  \bibinfo {author} {\bibfnamefont {P.}~\bibnamefont {Wang}}, \ and\ \bibinfo
  {author} {\bibfnamefont {A.}~\bibnamefont {Barabasi}},\ }\href {\doibase
  10.1038/nphys1760} {\bibfield  {journal} {\bibinfo  {journal} {Nat Phys}\
  }\textbf {\bibinfo {volume} {6}},\ \bibinfo {pages} {818} (\bibinfo {year}
  {2010})}\BibitemShut {NoStop}%
\bibitem [{\citenamefont {Burov}\ \emph {et~al.}(2011)\citenamefont {Burov},
  \citenamefont {Jeon}, \citenamefont {Metzler},\ and\ \citenamefont
  {Barkai}}]{burov11}%
  \BibitemOpen
  \bibfield  {author} {\bibinfo {author} {\bibfnamefont {S.}~\bibnamefont
  {Burov}}, \bibinfo {author} {\bibfnamefont {J.}~\bibnamefont {Jeon}},
  \bibinfo {author} {\bibfnamefont {R.}~\bibnamefont {Metzler}}, \ and\
  \bibinfo {author} {\bibfnamefont {E.}~\bibnamefont {Barkai}},\ }\href
  {\doibase 10.1039/c0cp01879a} {\bibfield  {journal} {\bibinfo  {journal}
  {Physical Chemistry Chemical Physics}\ }\textbf {\bibinfo {volume} {13}},\
  \bibinfo {pages} {1800} (\bibinfo {year} {2011})}\BibitemShut {NoStop}%
\bibitem [{\citenamefont {Saxton}(2007)}]{saxton07}%
  \BibitemOpen
  \bibfield  {author} {\bibinfo {author} {\bibfnamefont {M.~J.}\ \bibnamefont
  {Saxton}},\ }\href {\doibase 10.1529/biophysj.106.092619} {\bibfield
  {journal} {\bibinfo  {journal} {Biophys. J.}\ }\textbf {\bibinfo {volume}
  {92}},\ \bibinfo {pages} {1178} (\bibinfo {year} {2007})}\BibitemShut
  {NoStop}%
\bibitem [{\citenamefont {Bouchaud}\ and\ \citenamefont
  {Georges}(1990)}]{bouchaud90}%
  \BibitemOpen
  \bibfield  {author} {\bibinfo {author} {\bibfnamefont {J.}~\bibnamefont
  {Bouchaud}}\ and\ \bibinfo {author} {\bibfnamefont {A.}~\bibnamefont
  {Georges}},\ }\href {\doibase 10.1016/0370-1573(90)90099-N} {\bibfield
  {journal} {\bibinfo  {journal} {Phys. Rep.}\ }\textbf {\bibinfo {volume}
  {195}},\ \bibinfo {pages} {127} (\bibinfo {year} {1990})}\BibitemShut
  {NoStop}%
\bibitem [{\citenamefont {Goychuk}\ and\ \citenamefont
  {H{\"a}nggi}(2002)}]{goychuk02}%
  \BibitemOpen
  \bibfield  {author} {\bibinfo {author} {\bibfnamefont {I.}~\bibnamefont
  {Goychuk}}\ and\ \bibinfo {author} {\bibfnamefont {P.}~\bibnamefont
  {H{\"a}nggi}},\ }\href@noop {} {\bibfield  {journal} {\bibinfo  {journal}
  {Proc. Natl. Acad. Sci. USA}\ }\textbf {\bibinfo {volume} {99}},\ \bibinfo
  {pages} {3552} (\bibinfo {year} {2002})}\BibitemShut {NoStop}%
\bibitem [{\citenamefont {Mantegna}\ and\ \citenamefont
  {Stanley}(1994)}]{mantegna94}%
  \BibitemOpen
  \bibfield  {author} {\bibinfo {author} {\bibfnamefont {R.~N.}\ \bibnamefont
  {Mantegna}}\ and\ \bibinfo {author} {\bibfnamefont {H.~E.}\ \bibnamefont
  {Stanley}},\ }\href {\doibase 10.1103/PhysRevLett.73.2946} {\bibfield
  {journal} {\bibinfo  {journal} {Phys. Rev. Lett.}\ }\textbf {\bibinfo
  {volume} {73}},\ \bibinfo {pages} {2946} (\bibinfo {year}
  {1994})}\BibitemShut {NoStop}%
\bibitem [{\citenamefont {Koponen}(1995)}]{koponen95}%
  \BibitemOpen
  \bibfield  {author} {\bibinfo {author} {\bibfnamefont {I.}~\bibnamefont
  {Koponen}},\ }\href {\doibase 10.1103/PhysRevE.52.1197} {\bibfield  {journal}
  {\bibinfo  {journal} {Phys. Rev. E}\ }\textbf {\bibinfo {volume} {52}},\
  \bibinfo {pages} {1197} (\bibinfo {year} {1995})}\BibitemShut {NoStop}%
\bibitem [{\citenamefont {Nakao}(2000)}]{nakao00}%
  \BibitemOpen
  \bibfield  {author} {\bibinfo {author} {\bibfnamefont {H.}~\bibnamefont
  {Nakao}},\ }\href {\doibase DOI: 10.1016/S0375-9601(00)00059-1} {\bibfield
  {journal} {\bibinfo  {journal} {Phys. Lett. A}\ }\textbf {\bibinfo {volume}
  {266}},\ \bibinfo {pages} {282 } (\bibinfo {year} {2000})}\BibitemShut
  {NoStop}%
\bibitem [{\citenamefont {Gajda}\ and\ \citenamefont
  {Magdziarz}(2010)}]{gajda10}%
  \BibitemOpen
  \bibfield  {author} {\bibinfo {author} {\bibfnamefont {J.}~\bibnamefont
  {Gajda}}\ and\ \bibinfo {author} {\bibfnamefont {M.}~\bibnamefont
  {Magdziarz}},\ }\href {\doibase 10.1103/PhysRevE.82.011117} {\bibfield
  {journal} {\bibinfo  {journal} {Phys. Rev. E}\ }\textbf {\bibinfo {volume}
  {82}},\ \bibinfo {pages} {011117} (\bibinfo {year} {2010})}\BibitemShut
  {NoStop}%
\bibitem [{\citenamefont {Feller}(1971)}]{feller71c17}%
  \BibitemOpen
  \bibfield  {author} {\bibinfo {author} {\bibfnamefont {W.}~\bibnamefont
  {Feller}},\ }\href@noop {} {\emph {\bibinfo {title} {An Introduction to
  Probability Theory and its Applications}}},\ \bibinfo {edition} {2nd}\ ed.,\
  Vol.~\bibinfo {volume} {II}\ (\bibinfo  {publisher} {Wiley, New
  York},\ \bibinfo {year} {1971})\ Chap.~\bibinfo {chapter} {17}\BibitemShut
  {NoStop}%
\bibitem [{\citenamefont {Burov}\ \emph {et~al.}(2010)\citenamefont {Burov},
  \citenamefont {Metzler},\ and\ \citenamefont {Barkai}}]{burov10}%
  \BibitemOpen
  \bibfield  {author} {\bibinfo {author} {\bibfnamefont {S.}~\bibnamefont
  {Burov}}, \bibinfo {author} {\bibfnamefont {R.}~\bibnamefont {Metzler}}, \
  and\ \bibinfo {author} {\bibfnamefont {E.}~\bibnamefont {Barkai}},\ }\href
  {\doibase 10.1073/pnas.1003693107} {\bibfield  {journal} {\bibinfo  {journal}
  {Proc. Natl. Acad. Sci. U.S.A}\ }\textbf {\bibinfo {volume} {107}},\ \bibinfo
  {pages} {13228 } (\bibinfo {year} {2010})}\BibitemShut {NoStop}%
\bibitem [{\citenamefont {Jeon}\ \emph {et~al.}(2011)\citenamefont {Jeon},
  \citenamefont {Tejedor}, \citenamefont {Burov}, \citenamefont {Barkai},
  \citenamefont {Selhuber-Unkel}, \citenamefont {Berg-S\o{}rensen},
  \citenamefont {Oddershede},\ and\ \citenamefont {Metzler}}]{jeon11}%
  \BibitemOpen
  \bibfield  {author} {\bibinfo {author} {\bibfnamefont {J.-H.}\ \bibnamefont
  {Jeon}}, \bibinfo {author} {\bibfnamefont {V.}~\bibnamefont {Tejedor}},
  \bibinfo {author} {\bibfnamefont {S.}~\bibnamefont {Burov}}, \bibinfo
  {author} {\bibfnamefont {E.}~\bibnamefont {Barkai}}, \bibinfo {author}
  {\bibfnamefont {C.}~\bibnamefont {Selhuber-Unkel}}, \bibinfo {author}
  {\bibfnamefont {K.}~\bibnamefont {Berg-S\o{}rensen}}, \bibinfo {author}
  {\bibfnamefont {L.}~\bibnamefont {Oddershede}}, \ and\ \bibinfo {author}
  {\bibfnamefont {R.}~\bibnamefont {Metzler}},\ }\href {\doibase
  10.1103/PhysRevLett.106.048103} {\bibfield  {journal} {\bibinfo  {journal}
  {Phys. Rev. Lett.}\ }\textbf {\bibinfo {volume} {106}},\ \bibinfo {pages}
  {048103} (\bibinfo {year} {2011})}\BibitemShut {NoStop}%
\bibitem [{\citenamefont {Rebenshtok}\ and\ \citenamefont
  {Barkai}(2007)}]{rebenshtok07}%
  \BibitemOpen
  \bibfield  {author} {\bibinfo {author} {\bibfnamefont {A.}~\bibnamefont
  {Rebenshtok}}\ and\ \bibinfo {author} {\bibfnamefont {E.}~\bibnamefont
  {Barkai}},\ }\href {\doibase 10.1103/PhysRevLett.99.210601} {\bibfield
  {journal} {\bibinfo  {journal} {Phys. Rev. Lett.}\ }\textbf {\bibinfo
  {volume} {99}},\ \bibinfo {pages} {210601} (\bibinfo {year}
  {2007})}\BibitemShut {NoStop}%
\bibitem [{\citenamefont {Rebenshtok}\ and\ \citenamefont
  {Barkai}(2008)}]{rebenshtok08}%
  \BibitemOpen
  \bibfield  {author} {\bibinfo {author} {\bibfnamefont {A.}~\bibnamefont
  {Rebenshtok}}\ and\ \bibinfo {author} {\bibfnamefont {E.}~\bibnamefont
  {Barkai}},\ }\href {http://dx.doi.org/10.1007/s10955-008-9610-3} {\bibfield
  {journal} {\bibinfo  {journal} {J. Stat. Phys.}\ }\textbf {\bibinfo {volume}
  {133}},\ \bibinfo {pages} {565} (\bibinfo {year} {2008})}\BibitemShut
  {NoStop}%
\bibitem [{\citenamefont {Akimoto}(2008)}]{akimoto08b}%
  \BibitemOpen
  \bibfield  {author} {\bibinfo {author} {\bibfnamefont {T.}~\bibnamefont
  {Akimoto}},\ }\href@noop {} {\bibfield  {journal} {\bibinfo  {journal}
  {{\it ibid.}}\ }\textbf {\bibinfo {volume} {132}},\ \bibinfo {pages} {171}
  (\bibinfo {year} {2008})}\BibitemShut {NoStop}%
\end{thebibliography}


%section {figures}
\end{document}